\documentclass[12pt]{article}
\usepackage{graphicx} 
\usepackage{url}
\usepackage[dvips]{hyperref}
\usepackage{breakurl}
\usepackage{csquotes}
\usepackage{tikz}
\usetikzlibrary{shapes,arrows}
\usepackage{microtype}

\usepackage[
backend=biber,
style=footnote-dw,
sorting=none,
hyperref=true,
backref=true,
]{biblatex}

\title{The Manhattan Trap: \\ \large Why a Race to Artificial Superintelligence is Self-Defeating}
\date{December 2024}
\author{
  Corin Katzke\footnote{Author correspondence: corin.katzke@convergenceanalysis.org}\\
  \small{Convergence Analysis}\\
  \and 
  Gideon Futerman\\
  \small{University of Oxford}\\
}
\begin{document}

\maketitle{}

\begin{abstract}
This paper examines the strategic dynamics of international competition to develop Artificial Superintelligence (ASI). We argue that the same assumptions that might motivate the US to race to develop ASI also imply that such a race is extremely dangerous. These assumptions—that ASI would provide a decisive military advantage and that states are rational actors prioritizing survival—imply that a race would heighten three critical risks: great power conflict, loss of control of ASI systems, and the undermining of liberal democracy. Our analysis shows that ASI presents a \textit{trust dilemma} rather than a \textit{prisoner's dilemma}, suggesting that international cooperation to control ASI development is both preferable and strategically sound. We conclude that cooperation is achievable. 
\end{abstract}

\newpage
\tableofcontents

\newpage
\section*{Executive Summary}
\label{sec: exec}

This paper examines the strategic dynamics of international competition to develop Artificial Superintelligence (ASI). We argue that the same assumptions that might motivate the US to race to develop ASI also imply that such a race is extremely dangerous. 

A race to develop ASI is motivated by two assumptions: that ASI provides a decisive military advantage (DMA) to the first state that develops it, and that states are rational actors aware of ASI's strategic implications.

However, these same assumptions make racing catastrophically dangerous for three reasons.

First, an ASI race creates a threat to strategic stability that could trigger a war between the US and its adversaries, particularly China. If ASI could provide a decisive military advantage, states would rationally view their adversaries' ASI development as an existential threat justifying military intervention and espionage. A state cannot race to ASI without incurring significant risk of great power conflict unless either (a) its adversaries are unaware of ASI's importance (contradicting the awareness assumption), (b) development can be kept secret (likely impossible given the scale required), or (c) adversaries believe they can win the race (requiring a close competition in which victory is not assured).

Second, racing heightens the risk of losing control of an ASI system once developed. We do not take a position on the likelihood of loss of control. Instead, we observe that the argument for racing assumes that ASI would wield a decisive advantage over the militaries of global superpowers; accordingly, losing control of such a technology would also present an existential threat to the state that developed it. We also argue that an ASI would only provide a DMA if its capabilities scaled extremely rapidly—precisely the scenario in which loss of control risk is theoretically highest. Finally, a race compounds this risk by creating competitive pressure to develop ASI before adequate control measures are in place. 

Third, even a controlled ASI threatens to disrupt the internal power structures of the state that develops it. In the case of a US, a successful ASI project would likely undermine the liberal democracy it purports to defend. An ASI system that would grant a state international advantage would also grant its controllers unprecedented domestic power, creating an extreme concentration of power likely incompatible with democratic checks and balances. Racing exacerbates this risk by requiring development to occur quickly and in secret—and therefore without public input.

These three dangers—great power conflict, loss of control, and power concentration—represent three successive barriers a state would need to overcome to `win' a race to ASI. Together, they imply that an ASI race poses an existential threat to the national security of the states involved. Assuming states are informed and rational, the strategic situation can be modeled as a \textit{trust dilemma}: states would prefer mutual restraint to racing and will only race if they believe others will. States can therefore avoid an ASI race by establishing a verification regime—a set of mechanisms that can verify a state's compliance with an international agreement not to pursue an ASI project. An ASI project would be highly distinguishable from civilian AI applications and not integrated with a state's economy—precisely the conditions under which verification regimes have historically succeeded.

The paper concludes that, if an ASI race is motivated, then cooperation to avoid an ASI race is both preferable and strategically sound. The assumptions that make racing seem necessary are the very reasons it is unwise.

\newpage
\section{Introduction}
\label{sec: intro}

In the coming years, it may become possible to develop Artificial Superintelligence (ASI)—an AI system that vastly exceeds human performance across nearly all cognitive tasks.\footcite{bostrom2014superintelligence} If so, the resources required by the first ASI project would limit potential developers to well-resourced states in control of AI infrastructure like AI labs, datacenters, and chip fabrication plants. The US—in collaboration with allies—would therefore be uniquely placed to develop ASI;\footcite{okeefe2024chips} however, China may also be able to competitively pursue such a project.\footcite{leopold_situational_2024} 

ASI has the potential to determine the future of state power. The defining strategic choice facing the United States is whether to pursue unilateral advantage through a US-led project or establish a multilateral control regime that includes China.\footcite{zelikow2024defense} Recent writing has outlined two starkly different views regarding this choice.

The first view advocates for a US-led effort to develop ASI analogous to the Manhattan Project. \footcite{leopold_situational_2024}\footcite{amodei_machines_2024}\footcite{altman_who_2024} A coalition of democracies would develop ASI first, and use this advantage to protect liberal democracy from ASI-enabled authoritarianism. Advocates argue that, given ASI's potential to confer a decisive military advantage, an ASI race is inevitable and attempts at international cooperation may be dangerously naive. 

This view is currently gaining influence in US politics. For example, the U.S.-China Economic and Security Review Commission’s 2024 report recommended “Congress establish and fund a Manhattan Project-like program dedicated to racing to and acquiring an Artificial General Intelligence (AGI) capability.”\footcite{noauthor_2024_2024} However, it has not yet significantly influenced US military and industrial policy.\footcite{institute_us-china_2023} The dominant view in US politics remains that the US should preserve a general technological advantage over China. This position is distinct from advocating for an ASI-specific race, and, as we will argue, pursuing general technological advantage need not include a race to ASI. 

The second view argues that ASI development is extremely dangerous, and international cooperation is necessary to avoid these dangers.\footcite{kissinger_henry_2023}\footcite{bengio_implications_2024}\footcite{aguirre_close_2024} International cooperation including China could avoid an ASI race and protect against AI-enabled threats by establishing safety standards, verification regimes, and compute controls. Some have even suggested that this cooperation ought to go as far as a single international ASI project, or a global moratorium on ASI development.\footcite{Hausenloy2023-ek}\footcite{MiottiUnknown-ye}

This strategic tension—racing to ASI or cooperating to reduce risks—is the central question of ASI geopolitics. The stakes could not be higher: superintelligent AI systems may determine not just the balance of power between states, but also the survival and freedom of humanity. How the United States navigates this dilemma in the coming years will likely prove decisive.

\section{Assumptions Motivating an ASI Race}
\label{sec: assumptions}

The future of AI development is difficult to predict. The academic community remains divided on crucial uncertainties, or “strategic parameters,” such as timelines to ASI and the risk of loss of control of ASI.\footcite{katzke2024scenario} These significantly impact how states should approach the implications of AI development on national security.

We do not attempt to resolve these uncertainties. Instead, we observe that the argument that the US should race to develop ASI to avoid being outpaced by competitors requires several assumptions. We then make an internal critique of that argument: under the assumptions that motivate an ASI race, such a race is extremely dangerous. We conclude that, given those dangers, cooperation is strategically preferable.

\subsection{Decisive Military Advantage}
\label{subsec: dma}

The argument for racing to develop ASI assumes that ASI will provide a decisive military advantage (DMA) over states without ASI. Former OpenAI researcher Leopold Aschenbrenner makes this assumption explicit, writing that ASI "will give a decisive military advantage, perhaps comparable only with nuclear weapons [...] those who have it will wield complete dominance over those who don't."\footcite{leopold_situational_2024} Similarly, the CEO of Anthropic, Dario Amodei, calls for a coalition of democracies to "use AI to achieve robust military superiority."\footcite{amodei_machines_2024} If ASI provides a DMA, the racing argument proceeds that it would be intolerable for the US to be put at a decisive military disadvantage by an adversary who develops ASI first. Therefore, to protect itself, the US must develop ASI first. 

We use the term 'ASI' as the object of our analysis because it proves the clearest case for obtaining a decisive military advantage. Other possible (and mutually inclusive) terms include artificial general intelligence (AGI) and Transformative AI (TAI)—AI that precipitates a transition comparable to the agricultural or industrial revolution.\footcite{holden_background_2016} In the case of AGI, a state might gain advantage as the result of diffusing AGI through its economy, and not as the result of leading in AGI development. Moreover, it may be plausible for multiple superhuman narrow AI systems to bring similar economic benefits to AGI. Similarly, if TAI development followed a trajectory like the Industrial Revolution—with technologies diffusing gradually through economies—then the state leading AI development would not necessarily gain a decisive military advantage. Other states would likely be able to maintain relative military parity with their own development and diffusion.\footcite{ding_engines_2023}

ASI could provide a decisive military advantage to its first developer given a few conditions. First, ASI must create an extremely rapid or discontinuous leap in military capability rather than a gradual improvement in military capability. Gradual capability improvement could result in a DMA if a single state developed ASI unimpeded—however, this is unlikely. Advantage is necessarily relative. If the capability improvement conferred by ASI development was gradual, then other states could develop their own ASI projects and maintain relative parity. 

Additionally, if it is clear to states early on that an ASI project will eventually result in a DMA, then they can interfere in that project. Military intervention or coercion could halt ASI development before it realizes a DMA. The racing argument requires ASI to be more like nuclear weapons: a singular development that creates an immediate and dramatic shift in the balance of power.

Second, ASI must be able to undermine existing military deterrents, particularly nuclear deterrents. For ASI to provide a decisive military advantage, it would need to undermine the ability of a nuclear power to credibly threaten nuclear retaliation.\footcite{leopold_situational_2024} 

\subsection{Defensive Realism}
\label{subsec: dr}

The argument for racing to develop ASI also requires the assumption that states are (to some extent) rational actors and are well-informed about ASI’s strategic implications. If a state was not rational or well-informed, then it would not necessarily be motivated to race to ASI.

First, our analysis will model states as rational actors acting according to defensive realism—that is, states prioritize their survival and security, acting primarily to prevent threats to their survival.\footcite{jervis_realism_1999} Advocates for racing to ASI argue that the US must do so to prevent adversaries from gaining decisive military advantage through ASI first—not out of a pursuit of expansion or hegemony. Second, the racing argument assumes states are well-informed about the possibility of ASI providing a decisive military advantage. If a state lacked that awareness, then it would lack a motivation to race to ASI.

For the remainder of this paper, we assume (but do not argue for) these conditions about ASI development and state behavior in order to mount an internal critique of the racing argument.

\section{Dangers of an ASI Race}
\label{sec: dangers}

In this section, we establish three reasons why, given the assumptions above, racing to develop ASI is dangerous. First, a state ASI project is likely to incite adversaries to threaten or undertake military intervention to prevent its completion, which could precipitate great power conflict. Second, the rate of development required to establish a decisive military advantage creates a higher risk of losing control of ASI. Finally, even a controlled ASI threatens extreme power concentration and the disruption of a state’s internal political organization—in the case of the US, the end of liberal democracy. These dangers imply that states—and the US in particular—should prefer slow and cooperative development. 

\subsection{Great Power Conflict}
\label{subsec: conflict}

Arms races between states can cause conflict.\footcite{gibler_taking_2005} Arms development promises to improve a state’s military power relative to an adversary; in order to maintain its relative power, then, that adversary’s arms development must keep pace. One cause of conflict arises when a state can’t keep pace: in that case, it might try to disrupt its adversary’s arms development. This may be particularly likely if a state’s adversary is on track to gain a decisive military advantage, since this poses an existential threat to its security.

The racing argument assumes that ASI grants a decisive military advantage to the first state that develops it. This suggests there are significant analogies between ASI races and arms races. There are several ways of characterizing this decisive military advantage, but perhaps the clearest impetus for conflict is that such an ASI would undermine a state’s nuclear deterrent.\footcite{leopold_situational_2024} In the age of nuclear powers, a state’s security is preserved by credibly committing to nuclear retaliation in the case of attack by an adversary.\footcite{schelling_strategy_1960} Even if the attacker might ultimately prevail, it is deterred by the destruction a nuclear retaliation would cause. For example, the US might ultimately prevail in an all-out war with China—but it is not willing to risk, say, the destruction of a major American city that such a war might cause. Nuclear deterrence (between a small number of states) therefore preserves strategic stability. 

ASI threatens to undermine the strategic stability between nuclear states. The only formal agreement on strategic stability to date is the \textit{Soviet-United States Joint Statement on Future Negotiations on Nuclear and Space Arms and Further Enhancing Strategic Stability}.\footcite{bush_soviet-united_1990} While this agreement was made between the USA and the Soviet Union, it provides the basis for a contemporary understanding of strategic stability. It attempted to eliminate the “incentives for a first nuclear strike” by “giving priority to [the creation and maintenance of] highly survivable systems,” which could not be reliably destroyed in a first strike. However, the development of ASI undermines the survivability of an adversary state’s nuclear retaliation systems. If ASI could plausibly break deterrence, then a US ASI project would rationally be seen as an existential threat by a nuclear-armed US adversary—in particular, China and Russia. This might warrant a preemptive attack on US ASI development infrastructure, such as data centers and semiconductor fabrication plants, in order to prevent the project’s completion. 

Advocates for an ASI project argument give the Chinese and Russian states clear reason to be worried for their security. For example, Amodei explicitly argues that ASI ought to be used to achieve “robust military superiority” leading to an “eternal 1991.”\footcite{amodei_machines_2024} If the US project is perceived as an existential threat to its adversaries, these adversaries might risk war in order to stop a US ASI project. This would be a rational response for states concerned with maintaining their independence and security.  

Of course, eliciting a full-scale nuclear retaliation from the US would pose a similar threat to allowing the US to develop ASI. However, an adversary might judge that a limited preemptive attack could be contained to a limited war. A limited attack might be carried out by cyber offensives, conventional weapons, or even tactical nuclear weapons. Indeed, the idea of limited nuclear strikes has recently been suggested by Russian military planners.\footcite{alberque_russian_2024} 

Given both sides would want to avoid nuclear conflict, the risk of a US nuclear retaliation would not be certain. If an adversary thought it could destroy enough US ASI development infrastructure in a limited attack to remove the US’ lead, then that risk of escalation might be preferable to allowing the US to develop ASI and permanently undermine its security. These attacks would necessarily be preemptive—that is, before the ASI has been developed. As we discussed above, if deterrence is possible after ASI has been developed then ASI does not create a decisive military advantage. 

While China maintains an official no-first-use policy for nuclear weapons, China's recent nuclear buildup has cast some doubt on this policy.\footcite{noauthor_2021_2021} Moreover, a preemptive strike could involve conventional weapons or cyber warfare. The possibility of a non-nuclear attack might actually increase the likelihood of a preemptive strike, as it could be perceived as less likely to provoke full-scale US retaliation.

Similarly to the agreement on strategic stability between the US and USSR, Chinese strategists emphasize the concept of 'asymmetric strategic stability,' which relies on maintaining mutual vulnerability between powers.\footcite{kaufman_prc_2023}\footcite{perkovich_engaging_2022} ASI threatens to eliminate this vulnerability entirely, for the state that controls ASI destabilizes the current balance and permanently undermines Chinese security. 

Chinese policy, as with many states' nuclear weapons policies, is difficult to predict—particularly in the context of a race to ASI. Nonetheless, alongside reasons for why preemptive strikes might be rationally favored for states defending their national security, there are other lines of evidence for this outcome. For example, one war game modeling a race to ASI regularly observes ‘runners-up’ engaging in direct military conflict.\footcite{gruetzemacher_strategic_2024} The state behind in ASI development “will likely feel justified in launching such attacks because, without a preexisting treaty regarding other states’ autonomy post-RTAI [radically-transformative AI], the deployment of RTAI could be perceived as an existential threat to states not allied with the deployer.”

Instead of preemptive military intervention to hinder a US ASI project, the US’ adversaries might also attempt to keep pace with a US project by stealing its research and outputs. In particular, the risk of model weight exfiltration poses a significant threat to maintaining a US lead.\footcite{nevo_securing_2024} Sophisticated state actors may be capable of stealing critical data despite stringent security measures. Such exfiltration could allow rivals to rapidly close technological gaps at minimal cost, effectively negating some of the lead gained through extensive investment and research.

Proposed extreme security measures, such as air-gapping and intense surveillance, come with their own set of challenges. These measures may incur substantial costs, potentially deterring top talent from joining the project. Furthermore, the extreme centralization that such security measures likely entail would make a state’s ASI development more vulnerable to a single military attack by an adversary. 

Perfect security is impossible. Human intelligence remains a persistent vulnerability, whether through human error, ideological motivations, bribery, or extortion. For example, even the Manhattan Project leaked key information to the USSR.\footcite{noauthor_espionage_nodate} The dynamic nature of AI scientists' views on ASI development adds another layer of unpredictability to the security equation. A US ASI project might inadvertently support its adversaries' ASI development through exfiltrated research, turning their efforts into a self-defeating endeavor. 

A US ASI projects therefore creates a dilemma: if espionage is possible, then the US' adversaries may keep pace with US ASI development. If espionage is not possible, then the project may precipitate military intervention.

In order to develop ASI without creating a significant risk of great power conflict, one of three conditions would need to hold. First, the US’ nuclear-armed adversaries would need to have not “woken up” to the possibility or implications of ASI. If this were true, however, then the US would not have a reason to race to ASI, since the motivation for a US ASI project depends on the assumption that its adversaries are racing to develop ASI.

Second, the US would need to be able to mislead its adversaries as to the state of its ASI development. This would require very high levels of secrecy. Given the difficulties of even securing model weights against state actors (Section 3.1.3), as well as the possibilities of monitoring energy usage, infrastructure, supply chains, and researcher activity involved in an ASI project, it seems unlikely that China wouldn’t be able to realize that the US is near to ASI. For the same reasons that a verification regime is promising (Section 5), keeping an ASI project secret is unlikely.

The final justification would be that China would perceive itself as likely enough to win the race to ASI so as to not risk an attack on the US. This requires a dangerously close race where almost all AI development resources are put towards creating ASI. The safety-buffer that is often cited as a justification for fast ASI development would be completely absent.

Acknowledging the risk of great power conflict does not mean the US must abandon its national interests. However, the danger of conflict associated with an ASI race suggests that mutual restraint is in the best interests of all states involved. Cooperation on ASI development limitations could become a targeted area of agreement between the US and its adversaries to avoid conflict.

\subsection{Loss of Control}
\label{subsec: control}

The dangers of an ASI race would not be averted even if a US project was able to avoid or overcome interstate conflict. The argument for an ASI race assumes that ASI would wield a decisive advantage over the militaries of global superpowers. Accordingly, it would be the most potent technology ever developed. 

ASI also carries a risk unprecedented in technological development: loss of control. While the nature of other powerful technologies like nuclear weapons requires “humans in the loop,” ASI would be able to act autonomously— indeed, autonomous action is a key capability involved in ASI’s military potential. “Loss of control” refers to the possibility that we lose the ability to meaningfully constrain an ASI’s behavior, even if it began to act in ways we did not intend.\footcite{noauthor_international_2024}

Experts are divided on the difficulty of controlling an ASI, but many AI scientists, AI companies, and policymakers see loss of control as a risk to be taken extremely seriously.\footcite{noauthor_statement_nodate} In order to provide a DMA, ASI systems will likely be designed as goal-directed agents able to act autonomously. To control such a system, we would need to adequately specify which goals we want these systems to pursue.\footcite{bostrom2014superintelligence} Even if we do train an ASI on adequately specified goals, it may still fail to form an accurate internal representation of these goals.\footcite{Langosco2021-bw}\footcite{Ngo2022-sc} Even seemingly innocuous unintended goals may lead to loss of control, as a sufficiently intelligent agent might seek power instrumentally for a wide array of possible goals.\footcite{Carlsmith2023-ty}\footcite{katzke2024investigating} Without consensus that ASI will be well-controlled, its development represents an extreme case of “technology roulette.”\footcite{danzig_technology_2018} Losing control over a system more capable than national militaries would put the state that developed it—and all humanity—in a state of extreme vulnerability.

We do not take a position on the likelihood of loss of control. Instead, we argue this risk is much higher with the assumption that ASI will grant a DMA than without it. Risk involves two components: the severity and likelihood of harm. The DMA assumption first affects the risk involved in developing ASI by maximizing the severity of loss of control. A system capable of overwhelming global military powers would, by definition, pose an existential threat to the security of the state that developed it if it became uncontrolled. 

The DMA assumption also makes loss of control more likely. An ASI with a DMA would not only need to have a decisive advantage over current militaries, but also over other frontier AI systems (and all plausible coalitions of militaries and AI systems) at the time of its development. The DMA assumption therefore suggests that an extremely rapid pace of improvement in AI capabilities around the level of ASI is more likely.\footnote{While it's possible for a DMA to result from a frontrunner 'pulling away' due to gradual but sustained returns on ASI development rather than rapid capabilities improvements, this path to a DMA is less likely for two reasons. First, the result of the race would be clear to losing states in advance, such that military intervention or espionage could be employed far before the DMA is generated. Second, gradual returns on development are unlikely to be contained within a single state, but instead diffuse across state boundaries. If ASI development is gradual, then returns would likely follow a trajectory similar to the GPT diffusion model,(\footcite{ding_engines_2023}) which suggests that a state does not need to be at the technological forefront in order to remain competitive. In that case, a frontrunner likely wouldn't ever "pull away" enough to generate a DMA.}  This might occur if automated AI R\&D creates a feedback loop, or a single system recursively self-improves.\footcite{owen_automation_2024} Such a pace makes the development of technical control methods extremely difficult: control methods may not scale across orders of magnitude, and the time to develop new methods would be extremely short. The competitive pressures involved in an ASI race make this problem even more difficult, as a race incentivizes states to reallocate resources away from control research and towards capabilities research.\footcite{armstrong_racing_2016}\footcite{gruetzemacher_strategic_2024} States might risk quicker but more dangerous development pathways to ASI. 

In summary, the assumption that motivates an ASI race—that it would grant a decisive military advantage—implies that the probability of losing control of an ASI system is high and the consequences catastrophic.

\subsection{Power Concentration}
\label{subsec: power}

Even a controlled ASI poses dangers that undermine the motivation for an ASI race. The most common motivation for the argument that the US needs to beat China in a race to ASI is that a Chinese ASI would lead to the victory of authoritarianism over liberal democracy.\footcite{leopold_situational_2024} Defending liberal democracy ought to be a core US priority—however, the Chinese state winning a race to ASI is not the only way for liberal democracy to give way to authoritarianism. A successful US ASI project also poses an existential threat to liberal democracy.

Advocates for a US ASI project have cited the liberal, democratic, and republican character of the US as the normative grounding for why a world dominated by the US is more desirable than one dominated by China or Russia. Assuming the US maintains this character, we agree. The extreme concentrations of power accumulated by Xi Jinping and Vladimir Putin indeed highlight the danger of a lack of checks and balances in government. Aschenbrenner hopes “that we can instead rely on the wisdom of the Framers—letting radically different values flourish, and preserving the raucous plurality that has defined the American experiment.”\footcite{leopold_situational_2024} Similarly, the \textit{Special Competitive Studies Project} stresses that “the Founding Fathers, wary of overly concentrated power in a single institution or individual, designed a government where the legislative, executive, and judicial branches share and constrain each other’s authority.”\footcite{The-Special-Competitive-Studies-Project2024-bv} 

However, the development of an ASI system by the US, or actors in the US, could represent a similarly extreme concentration of power. The racing argument assumes that the state controlling such a system gains a decisive military advantage over other states—however, this advantage doesn’t just impact the balance of power internationally. Domestically, the small number of actors who would control an ASI—the one company, project leader, or part of the US national security establishment—would have a decisive advantage over the rest of the US, making, for example, a successful coup far more likely. 

Such a scenario represents a catastrophic failure of checks and balances in government, and the preservation of liberal democracy would rely on the goodwill and competence of the few actors that control ASI. This problem points to the fact that ensuring the development of ASI goes well is much more complicated than just technical control. For the same reason that we might want soldiers to refuse illegal orders, an ASI would need to adhere to moral or constitutional principles that supersede the instructions of its operators. The governance of ASI is a sociotechnical\footcite{weidinger2023sociotechnical} problem, calling for work not only in technical research, but also in systems design, political philosophy, and applied ethics. This work may be infeasible in a race scenario that reduces the time and transparency necessary for adequate political oversight. 

Ultimately, an ASI that preserves liberal democracy would have to be accountable to democratic processes and institutions. However, there are reasons to think ASI may be incompatible with these processes and institutions.\footcite{deudney_bounding_2024} Liberal democracy emerged out of a tradition of republicanism that has successfully bounded despotic power in the defense of the public interest. ASI undermines republicanism by creating a source of power so significant that effective bounds seem incredible. If no effective bounds were possible, the best case scenario may be ‘enlightened despotism’—however, such a result is anathema to the “wisdom of the Framers.” Those who frame a race to ASI as protecting liberal democracy against authoritarianism are misguided if there is no way to make an ASI compatible with democracy. 

Even if it were possible to design ASI in a way compatible with liberal democracy, an ASI developed during a race would be the result of an unaccountable process. It would necessarily be developed quickly, in secret, and without public input. It is unlikely, even if the US ‘wins’ an ASI race, that the resulting distribution of power preserves liberal democracy. Therefore, racing to develop ASI undermines the normative authority the US would have for doing so.

\section{The ASI Trust Dilemma}
\label{sec: trust dilemma}

In \autoref{sec: dangers}, we suggest that racing to develop ASI is extremely dangerous. A race is only strategically motivated if ASI provides a decisive military advantage and states are aware of the potential for ASI to provide such an advantage—precisely the scenario where risks of conflict, loss of control, and the undermining of liberal democracy are highest. By default, then, states would be better off cooperating than racing to develop ASI. 

However, one might still argue that racing is an unfortunate but inevitable outcome: an ASI race might be a prisoner's dilemma in which the best strategy for all players leads to a collectively bad outcome. This section argues that this is incorrect: given its dangers, an ASI race is better modeled as a \textit{trust dilemma} than as a \textit{prisoner's dilemma}. In a trust dilemma, cooperation is not only preferable but strategically sound. 

\subsection{Modeling an ASI Race}
\label{subsec: modeling}

One way we might model the choice to race or cooperate follows a “rationalist” or “realist” model of international relations.\footcite{waltz_theory_1979} According to rationalism, states can be modeled as rational agents with utility functions. (Defensive) realism gives some content to this utility function: above all, states act in order to preserve their survival. A condition in which a state is put at the total mercy of an adversary is evaluated by such a utility function as a worst-case scenario.

Of course, an outcome in which all states lose—such as human extinction—is also a worst-case scenario according to a realist utility function. For example, Nathan Sears argues that: 

\begin{quote}
   “Since there can be no states or nations without the continuation of humanity, an existential threat to humankind is, by logical extension, an existential threat to national survival. [...] The failure of states to eliminate existential threats to humankind poses an empirical challenge—or exposes a theoretical inconsistency—to this core premise of realism.”\footcite{sears_great_2023}
\end{quote}

Sears’ observation is correct; however, his conclusion does not necessarily follow: a state may judge that eliminating an existential threat to humanity exposes it to a more likely threat to its national survival. 

Suppose, for example, that the US is on the precipice of developing ASI. It believes with 90\% probability that it will be able to adequately control ASI given existing technical safety techniques—correspondingly, it gives a 10\% chance that developing ASI will result in loss of control that threatens human survival. However, if the US does not develop ASI, then it believes with 20\% certainty that China will, which will either result in loss of control or place China in a position of decisive military advantage—which are evaluated equally. A realist model of national behavior therefore predicts that the US will choose to develop ASI despite its existential threat to humanity.

Of course, from the perspective of humanity, human survival is more important than preserving US national security. One could take the extreme position of the physicist Arthur Compton, who, when discussing the possibility that the first atomic test would vaporize the earth’s atmosphere, proclaimed: “Better to accept the slavery of the Nazis than to run the chance of drawing the final curtain on mankind!”\footcite{buck_bombend_1959} 

We, however, do not need to take this perspective. The possibility of cooperation does not depend on states taking the perspective of humanity—rather, it lies in the fact that, if there is a non-negligible probability that ASI development precipitates war, creates an uncontrolled ASI, or disrupts a states internal political system, then the US and China would both prefer not to develop ASI. Given the extreme dangers of an ASI race (Section 3), this preference is similarly extreme—although our analysis still holds even if this is only a mild preference. The crucial implication is that if one state believes that the other will not develop ASI, then that state will not develop ASI itself.

The strategic situation can be described in the language of game theory as a trust dilemma.\footcite{jervis_cooperation_1978} In a trust dilemma, both players would rather cooperate than defect; however, if one player defects, it is better for the other to defect as well. 
  
\begin{table}
    \centering
    \begin{tabular}{ccc}
         &  Cooperate& Defect\\
         Cooperate&  10,10& 2,0\\
         Defect&  0,2& 1,1\\
    \end{tabular}
    \caption{A Trust Dilemma}
    \label{tab:my_label}
\end{table}

This differs substantially from a prisoners' dilemma, where if one player cooperates, it is better for the other player to defect than to cooperate. This means the only Nash equilibrium in a prisoners' dilemma is for both states to defect, which is mutually destructive. In a trust dilemma, there are two Nash equilibria, one where both states cooperate, and one where they defect. States would both prefer to cooperate, so the challenge is to establish sufficient mutual confidence that both players will cooperate. For example, this might be accomplished by way of a conditional commitment device: player A credibly commits to cooperating conditional on player B cooperating. Such a commitment would be sufficient to induce player B to commit to cooperating.

An example of a relatively weak commitment device might involve a player making a public statement committing itself to cooperate. If it defects, then it risks undermining its credibility or facing other social consequences. A commitment device works if it alters a player's payoffs such that it is always in their best interest to cooperate even if the other player defects, making cooperation a dominant strategy. In that case, the other player will also be led to cooperate. However, the strength of a commitment device needs to match the stakes of the game. No realistic commitment between states will be able to change a game’s relative payoffs when the stakes are taken to be a state's survival. 

In the absence of a commitment mechanism, players might also establish mutual confidence of cooperation by way of compliance verification. For example, in the case of an arms control treaty, both states might agree to allow the other sufficient access to verify its compliance with the treaty. Verification differs from commitment in that it doesn’t change the payoffs of the game—both players would still defect if the other defected. However, neither has reason to defect, since defection would be visible and lead to mutual defection. 

\subsection{Navigating the Transparency-Security Tradeoff}
\label{subsec: tradeoff}

Arms race dynamics that can be modeled as trust dilemmas are relatively common between states. In general, arms development is costly because it consumes resources that might have otherwise been used for positive state goals. Coe and Vaynman calculate that the \$2.7 Trillion spent by the world’s militaries in 2016 could have otherwise been used to “end world poverty, provide bed nets to every person exposed to mosquito-borne disease, identify and treat every person whose infection with HIV would otherwise go undetected or untreated, and increase world spending on research and development by half.”\footcite{coe_why_2020} 

If arms development is costly, and verification is an effective device for arms treaties, then why are there not more successful arms treaties? This is the question investigated by Coe and Vaynman in their paper, \textit{Why Arms Control is So Rare}. Their answer: “The main impediment to arms control is the need for monitoring that renders a state’s arming transparent enough to assure its compliance but not so much as to threaten its security.”\footcite{coe_why_2020} They call this dynamic the transparency-security tradeoff.

An ideal verification process would be costless: it would reveal only whether a state is compliant with an arms control treaty, and nothing else. However, in practice, verification processes might inadvertently reveal sensitive information to the verifying state to which it otherwise wouldn’t have had access. For example, the process of verifying that a military is not employing a particular kind of weapon might reveal other information—such as which kinds of weapons it \textit{is} using.

Coe and Vaynman argue that the tradeoff between compliance transparency and protecting security-relevant information is a primary cause of the failure of arms control treaties. If compliance is insufficiently transparent, then a state will not be able to trust that the other is cooperating. On the other hand, if verification makes compliance so transparent as to threaten to reveal security-relevant information, then a state might prefer to accept the costs of an arms race. 

A related paper by Vaynman and Volpe suggests that the transparency-security tradeoff is determined by the dual-use nature of the technology being controlled. The authors argue that dual-use technology can be characterized along two dimensions: the distinguishability of its military and civilian uses, and the extent of its integration in a state’s military and civil society.\footcite{vaynman_dual_2023}

The less distinguishable a technology is, the greater transparency verification processes will require. The more integrated a technology is, the more likely verification processes will reveal security-relevant information. Therefore, arms control treaties are most likely to succeed when the technology being controlled is highly distinguishable and nonintegrated, and least likely to succeed when it is indistinguishable and highly integrated.

AI is a dual-use technology, which means that its development has both military and civilian applications. However, a race to ASI would still incur costs for the same reasons as other arms races. A race to ASI would make the allocation of resources within AI development less socially beneficial. Resources spent on military AI systems, including ASI, might draw resources away from safer civilian applications of AI, such as medical research or consumer products. More importantly, however, a race to ASI is costly because of the arguments made above: an ASI race would precipitate great power conflict; and, even if a state was able to develop ASI through such conflict, it would do so only with substantial risk to its survival and power structure given the current state of safety research and institutional design.

Arms control for AI in general is therefore unlikely to succeed. The military and civilian applications of general-purpose systems are nearly indistinguishable, and AI will likely see wide use across military and civilian society. 

However, the opposite may be true of ASI development control: ASI development would likely be distinguishable from most civilian  AI development, and, so long as it is not developed, unintegrated in a state's economy. For example, it seems likely that the information that states would risk revealing could be minimized with relatively minor efforts, such as avoiding co-locating ASI training clusters with inference compute used for AI in the military.\footnote{In the current paradigm of frontier AI development, an ASI project would require the very largest data centers, using chips that may be identified as being used for AI, which combined should be distinguishable from civilian AI development. However, if distributed training runs become feasible to develop ASI, or algorithmic improvements reduce the scale of the ASI training runs substantially, this may change, making ASI development control harder.} The prospects for ASI development control are promising. Additionally, since the relative benefit of a treaty for ASI is so large, states would likely accept a relatively high risk of revealing some sensitive information to comply with an ASI-specific treaty. This suggests that the transparency-security tradeoff does not need to be optimal for ASI development control to succeed.

\section{International Cooperation on ASI Development}
\label{sec: cooperation}

As we discussed in \autoref{subsec: tradeoff}, ASI development control is most promising if targeted to projects developing frontier general purpose systems, and not to AI systems more generally. States might also want to establish agreements regarding narrow systems with military applications—for example, autonomous military drones, systems used to automate cyberwarfare, or systems with the ability to develop novel bioweapons. However, since a treaty attempting to cover all AI development would be infeasible, it may be more appropriate to deal with these other systems under separate agreements.

In this section, we discuss the possibility of an ASI development control treaty. Some advocates of an ASI race recognize the costs of such a race; however, they believe that since cooperation is extremely unlikely, the second-best option is for the US to win. We argue that this thinking is confused, especially under realist assumptions about state behavior—it is precisely because a race to ASI is so costly that ASI development control is not only preferable but strategically sound.

\subsection{Verification}
\label{subsec: verification}

One reason advocates of an ASI race believe cooperation is unlikely is because they assume ASI development control would require extremely reliable verification mechanisms, and that such mechanisms are not feasible. Neither of these assumptions is supported.

Such advocates might be implicitly relying on a prisoner’s dilemma model of an ASI race: according to such a model, a state would take any opportunity to defect from an arms control agreement, since defecting is a dominant strategy. However, as \autoref{sec: trust dilemma} discussed, a better model of an ASI race is a trust dilemma. In a trust dilemma, both cooperate-cooperate and defect-defect are equilibria, and the former is preferable. Our analysis in \autoref{sec: dangers} suggests that the former is vastly preferable in the case of an ASI race. The only reason a state would defect from an ASI arms control agreement is if it believed another state was defecting. 

Of course, this choice is probabilistic: a state would have reason to defect if the \textit{probability} that another state was cooperating dipped below some critical level. However, the level of this critical probability lowers as the difference between the values of the ‘Cooperate’ and the ‘Defect’ equilibria widens. That level is always lower than certainty—therefore, the verification mechanism for an ASI control agreement does not need to be perfect. Since the values between the two equilibria are particularly wide due to the extreme dangers of an ASI race, verification could be far from perfect and still enable cooperation.

In fact, verification might play only a partial role in a state’s trust in an arms control agreement. If both parties to an agreement were rational actors, and viewed each other as rational actors, then both would understand that the other would have no reason to defect from such an agreement. It would be enough for both parties to demonstrate that they were (reliably) rational in order to establish trust in an arms control agreement. The primary role of a verification regime could be to establish an initial trust that would self-reinforce as states observed the treaty.\footnote{Of course, there may be other reasons to prefer a strong verification regime, such as preventing malicious use of AI by non-state actors.}Nonetheless, prospects for ASI development verification currently seem strong. An ASI project would be incredibly difficult to reliably keep secret. In the current paradigm of frontier AI development, an ASI project would require massive data center construction that would be easily observable by national technical means. This would not reveal information not already available to adversaries, and as a result not creating a transparency-security tradeoff. Beyond national technical means, other verification methods, such as whistleblowing, third-party auditors, and on-chip governance could form a “defense in depth,” collectively forming a robust verification regime.\footcite{wasil_verification_2024}\footcite{scher2024mechanisms}

Political will rather than technical means is the more important barrier to ASI development control. As Allan Krass writes in the context of nuclear arms control verification, "The most that verification can ever be is a tool to aid in the implementation of a process whose foundation is a mutually shared recognition of the futility and danger of the arms race and the will to act politically on this recognition."\footcite{Krass1985-ra}

\subsection{Enforcement}
\label{subsec: enforcement}

Some discussions of an international ASI control treaty assume that it will require extremely strong enforcement mechanisms, including military intervention to prevent defectors from completing ASI projects. For example, in an article for TIME, Eliezer Yudkowsky characterizes a successful agreement as follows: “If intelligence says that a country outside the agreement is building a GPU cluster, be less scared of a shooting conflict between nations than of the moratorium being violated; be willing to destroy a rogue datacenter by airstrike.”\footcite{yudkowsky2023pausing}

As we argued above, a country developing ASI would indeed likely cause conflict with its adversaries. However, an arms race is so mutually dangerous that, once such awareness was common among political leaders, the specter of such conflict may be enforcement enough. 

If the analysis of ASI development as a trust dilemma is correct, and states can be modeled as rational actors, then the need for enforcement of an ASI treaty would be unlikely. It would be in no state’s interest to develop ASI if they could trust to even a limited extent that other states were not developing ASI. That trust itself could be based on the perception that other states were themselves rational actors, and supported by the verification methods discussed in \autoref{subsec: verification}.

\section{Conclusion}
\label{sec: conclusion}

The strategic dynamics examined in this paper reveal a stark irony: the same assumptions that seem to make an ASI race necessary also make it catastrophically dangerous. If ASI provides a decisive military advantage, and states are rational and well-informed, then racing creates significant danger at each stage of development. The first danger is great power conflict: states would rationally view competitors' ASI projects as existential threats requiring military intervention. Second it the heightened risk of losing control of an ASI once developed: an ASI would only provide a decisive advantage if its capabilities scaled extremely rapidly—the scenario is which loss of control risk is greatest. Finally, even a controlled ASI threatens to concentrate power in ways that could destroy the very institutions—like liberal democracy—that an ASI project would aim to protect.

These dangers transform our understanding of the strategic situation. Rather than being trapped in an inevitable race, states face a trust dilemma in which mutual restraint is both preferable and achievable. The challenge now is to establish mutual understanding of the dangers of an ASI race, and international cooperation to prevent these dangers before they are realized. 

\vspace{12pt}

\textit{Acknowledgments: Justin Bullock, David Kristoffersson, and Elliot McKernon.}

\end{document}